\documentclass[reprint, amsmath, amssymb, aps,prl,superscriptaddress]{revtex4-1}

\usepackage{graphicx}
\usepackage{dcolumn}
\usepackage{bm}
\usepackage[normalem]{ulem}
\usepackage[colorlinks=true,linkcolor=red,citecolor=blue]{hyperref}
\usepackage{xcolor}

\newcommand{\lcf}{L}
\newcommand{\nv}{\hat{\bf n}}
\newcommand{\lmmin}{\log_{10}(M_{\rm min}/M_\odot h^{-1})}
\newcommand{\lmone}{\log_{10}(M_1/M_\odot h^{-1})}
\newcommand{\mnras}{Mon. Not. R. Astron. Soc.}
\newcommand{\jcap}{Journal of Cosmology and Astroparticle Physics}
\newcommand{\apjl}{Astrophysical Journal, Letters}
\newcommand{\apjs}{Astrophysical Journal, Supplement}
\newcommand{\aj}{Astronomical Journal}
\newcommand{\pasp}{Publications of the ASP}

\begin{document}


\title{First measurement of projected phase correlations and large-scale structure constraints}
\author{Felipe Oliveira Franco}
\email{felipe.oliveirafranco@physics.ox.ac.uk}
\affiliation{Department of Physics, University of Oxford, Denys Wilkinson Building, Keble Road, Oxford OX1 3RH, United Kingdom}
\author{Boryana Hadzhiyska}
\affiliation{Harvard-Smithsonian Center for Astrophysics, 60 Garden St., Cambridge, MA 02138, USA}
\author{David Alonso}
\affiliation{Department of Physics, University of Oxford, Denys Wilkinson Building, Keble Road, Oxford OX1 3RH, United Kingdom}

\date{\today}

\begin{abstract}
Phase correlations are an efficient way to extract astrophysical information that is largely independent from the power spectrum. We develop an estimator for the line correlation function (LCF) of projected fields, given by the correlation between the harmonic-space phases at three equidistant points on a great circle. We make a first, 6.5$\sigma$ measurement of phase correlations on data from the 2MPZ survey. Finally, we show that the LCF can significantly improve constraints on parameters describing the galaxy-halo connection that are typically degenerate using only two-point data.
\end{abstract}

\maketitle

\section{Introduction}\label{sec:intro}
The distribution of matter in the late-time Universe follows an intricate network of clusters, filaments, sheets, and voids known as the cosmic web. These structures are evident signatures of a strongly non-Gaussian field, and thus significant information regarding its origin and evolution should be encoded in correlators of order higher than 2. This has motivated the study of non-Gaussian observables, including higher-order correlators \citep{1977ApJ...217..385G,2001PhRvL..86.1434F,2000ApJ...544..597S,2017MNRAS.465.1757G,2102.01696}, marked correlations \citep{1609.08632,2001.11024}, counts in cells \citep{1911.11158}, nearest-neighbor statistics \citep{1907.00989,2007.13342}, peaks and voids \cite{0906.3512,2007.09013,2102.05049,2110.10135}, and other non-linear transformations of the density field \citep{2006.06298,2103.09247,2108.07821}. Nevertheless, the exploitation of non-Gaussian observables is still lagging behind standard analyses based on two-point correlation function (2PCFs). The challenge is three-fold: no optimal summary statistic exists for generic non-Gaussian fields in terms of data compression, the number of elements for general $N$-point correlators grows geometrically with $N$, posing a significant data analysis challenge, and often it is not possible to derive theoretical predictions for complex non-Gaussian observables, requiring the use of expensive simulation-based emulators. The ideal higher-order statistic should therefore address these three shortcomings: it should comprise a relatively small number of datapoints that are largely independent of the 2PCF, and for which a relatively simple theoretical prediction can be produced. Phase correlations (and in particular the line correlation function) are a possible avenue to address this problem.

As originally proposed by \cite{1211.5213,1409.3007}, the line correlation function (LCF) is the three-point correlation between the Fourier-space phases of the overdensity field at three equi-distant points lying on a straight line. The power spectrum is solely sensitive to the modulus of the Fourier coefficients, discarding the phase information that is irrelevant for Gaussian fields. Thus the field's phase will often contain a large fraction of the information discarded by the 2PCF \cite{astro-ph/0211408}. Phase maps furthermore enhance the signal from filamentary structures \cite{1211.5213,1409.3007}, a signature of non-Gaussianity. Thus, the line correlation function has been proposed as the simplest non-trivial phase correlator, although other configurations exist \cite{1903.11402}.

The LCF was originally proposed for application on three-dimensional datasets, such as spectroscopic surveys, and indeed it has been shown to be a powerful observable to constrain the growth of structure $f$ and the amplitude of perturbations $\sigma_8$ \cite{1705.04392,1805.10178,1806.10276,2005.06325}. However, photometric galaxy surveys probing the late-time structure through weak lensing and the projected clustering of galaxies, play a significant role in our current understanding of the Universe \citep{2105.13549}, and their impact will increase with the advent of Stage-IV surveys such as LSST \cite{2009arXiv0912.0201L}. Projected maps of the matter and galaxy distributions still preserve much of the underlying non-Gaussian structure. Therefore the use of higher-order statistics will be vital to optimize the scientific yield of these datasets. This motivates the study of phase correlations on the sphere. Although this idea has been explored in the context of weak lensing \cite{2109.08047}, its usefulness for projected galaxy clustering has not been studied, and no measurement on real data has been carried out to date. In this paper, we develop a simple estimator for the projected LCF of the galaxy distribution, apply it for the first time to existing data, and explore the additional constraining power it yields in combination with the 2PCF.

\section{Data}\label{sec:data}
The 2MASS Photometric Redshift catalog (2MPZ) \cite{1311.5246} is an almost all-sky, flux-limited galaxy sample extracted from the Two Micron All-Sky Survey Extended Source Catalogue (2MASS) \cite{astro-ph/0004318}, by cross-matching it with two additional all-sky data-sets: SuperCOSMOS XSC \cite{astro-ph/0108290,1607.01189} and the Wide-field Infrared Survey Explorer (WISE) \cite{1008.0031}. Photometric redshifts were estimated for all the sources common to the three catalogues using the ANNz algorithm \cite{astro-ph/0311058}. The resulting sample has a median redshift $\bar{z}=0.08$ and high photo-$z$ accuracy ($\sigma_{\delta z}\sim0.013$), thanks to the large photometric coverage and abundance of overlapping spectroscopy.

2MPZ is an ideal sample for a first estimate of the projected LCF. On the one hand, since it is photometric, it is not possible to carry out a robust analysis in 3D, and we must resort to projected statistics. On the other hand, given its low redshift and good photo-$z$ accuracy, the cosmic web filaments, to which phase correlations are particularly sensitive, are still clearly visible in projection.

To enhance the cosmic web features and the non-Gaussianity of the projected galaxy overdensity, we impose a cut in photo-$z$ space of $z_{\rm photo}\leq0.1$. To avoid systematic fluctuations in the galaxy overdensity caused by dust extinction and stars, we mask regions of the sky most dominated by Galactic reddening as quantified by \cite{astro-ph/9710327} (see \cite{1805.11525}). The resulting mask removes $\sim32\%$ of the sky, leaving 475{,}258 sources after the photo-$z$ cut. We use this sample to generate a map of the galaxy overdensity, given simply by $\delta_p=N_p/\bar{N}-1$, where $N_p$ is the number of galaxies in pixel $p$, and $\bar{N}$ is the mean number of galaxies per pixel in the unmasked region. We make use of the {\tt HEALPix} pixelization scheme \cite{astro-ph/0409513,2019JOSS....4.1298Z} with resolution parameter $N_{\rm side}=128$.

\section{The projected LCF}\label{ssec:meth.lcf}

For a scalar field $\delta$ on the sphere, the projected LCF $\lcf(\theta)$ is defined as the correlation between the harmonic-space phases at three equi-distant points lying on a great circle. I.e.
\begin{equation}
    \lcf(\theta)\equiv\left\langle \varepsilon({\sf R}^T_\theta\nv)\,\varepsilon(\nv)\,\varepsilon({\sf R}_\theta\nv)\right\rangle,
\end{equation}
    where $\nv$ is a unit vector on the sphere, and ${\sf R}_\theta$ is a rotation matrix satisfying $\nv^T{\sf R}_\theta\nv=\cos\theta$, but otherwise unconstrained. Here, $\varepsilon(\nv)$ is a map of the harmonic-space phase:
    \begin{align}\label{eq:phase_def}
      \varepsilon(\nv)\equiv\sum_{\ell m}Y_{\ell m}(\nv)\,\frac{\delta_{\ell m}}{|\delta_{\ell m}|},\hspace{6pt}
      \delta_{\ell m}=\int d^2\nv\,Y^*_{\ell m}(\nv)\delta(\nv).
    \end{align}
    As noted in \cite{1211.5213}, thus defined the amplitude of any phase correlation would depend on the smallest scale on which $\delta(\nv)$ has been sampled (e.g. on the choice of map pixelization). To avoid this, we will instead work with a smoothed version of the phase map. This corresponds to simply replacing
    \begin{equation}\label{eq:smoothing}
      \frac{\delta_{\ell m}}{|\delta_{\ell m}|}\rightarrow\frac{\delta_{\ell m}}{|\delta_{\ell m}|}w_\ell
    \end{equation}
    in Eq. \ref{eq:phase_def}, where $w_\ell$ is the harmonic transform of the smoothing kernel. In what follows we will use a Gaussian smoothing $w_\ell=\exp(-\ell(\ell+1)\sigma_\theta^2/2)$, where $\sigma_\theta$ is the smoothing scale in radians.

    In addition to the projected LCF, we will also make use of the power spectrum of the same field, simply defined as the harmonic-space variance of $\delta$: $C_\ell\equiv\langle|\delta_{\ell m}|^2\rangle$.

    As shown in \cite{Oliveira_inprep}, the projected LCF can be calculated from the angular bispectrum of $\varepsilon$, which itself is connected with the bispectrum and power spectrum of $\delta$ at leading order. The analytical derivation of this connection is described in detail in \cite{Oliveira_inprep}. Here, we will instead base our analysis on the use of an emulator, which replicates the measurement of $\lcf(\theta)$ on a set of simulated datasets. This is described in the next section.

    One of the main challenges to the use of the LCF in real data is the impact of an incomplete sky coverage. The calculation of the harmonic coefficients (and thence $\varepsilon_{\ell m}$) in the presence of a mask leads to non-trivial statistical couplings between different modes that could affect the shape and normalization of the resulting LCF. To build some intuition regarding the effects of a sky mask, consider the variance of $\varepsilon$ in the flat-sky approximation. In this case:
    \begin{eqnarray}
      \langle\varepsilon^2({\bf x})\rangle=\int\frac{d^2k\,d^2l}{(2\pi)^4}e^{i{\bf x}\cdot({\bf k}-{\bf q})}\frac{\delta_{\bf k}\delta_{\bf l}^*}{|\delta_{\bf k}||\delta_{\bf l}|}=A^{-1}\Omega_{\rm pix}^{-1},
    \end{eqnarray}
    where $A$ is the area of the flat-sky patch, and $\Omega_{\rm pix}$ is the discrete pixel size. To lowest order, the patch size therefore affects the overall normalization of $\varepsilon$ as $f_{\rm sky}^{-1/2}$, where $f_{\rm sky}$ is the sky fraction of the sample footprint. The effect of the pixel resolution is taken care of by the smoothing in Eq. \ref{eq:smoothing}.
    
    We therefore propose an approximate estimator for the LCF that simply corrects for this effect:
    \begin{equation}\label{eq:fsky}
      \hat{\lcf}(\theta)=f_{\rm sky}^{3/2}\tilde{\lcf}(\theta),
    \end{equation}
    where $\hat{\lcf}$ is our estimator and $\tilde{\lcf}$ is the LCF calculated via triplet counting on the masked $\varepsilon(\nv)$ map. Specifically, we compute $\tilde{\lcf}(\theta)$ as
    \begin{eqnarray}
      \tilde{\lcf}(\theta)=\frac{\sum_{ij}\varepsilon_i\varepsilon_j\varepsilon_{\bar{ij}}\Theta(\theta<\theta_{ij}/2<\theta+\Delta\theta)}{\sum_{ij}\Theta(\theta<\theta_{ij}/2<\theta+\Delta\theta)}.
    \end{eqnarray}
    Here, $i$ and $j$ run over all unmasked pixels, $\varepsilon_{i}$ is the value of the phase in pixel $i$, $\theta_{ij}$ is the angle between pixels $i$ and $j$, and $\varepsilon_{\bar{ij}}$ is the value of $\varepsilon$ in the pixel nearest to the mid-point between pixels $i$ and $j$ along the great circle connecting them. $\Theta$ is a top-hat function and $\Delta\theta$ is the width of the angular bins used in the analysis.
    \begin{figure}
      \centering
      \includegraphics[width=0.47\textwidth]{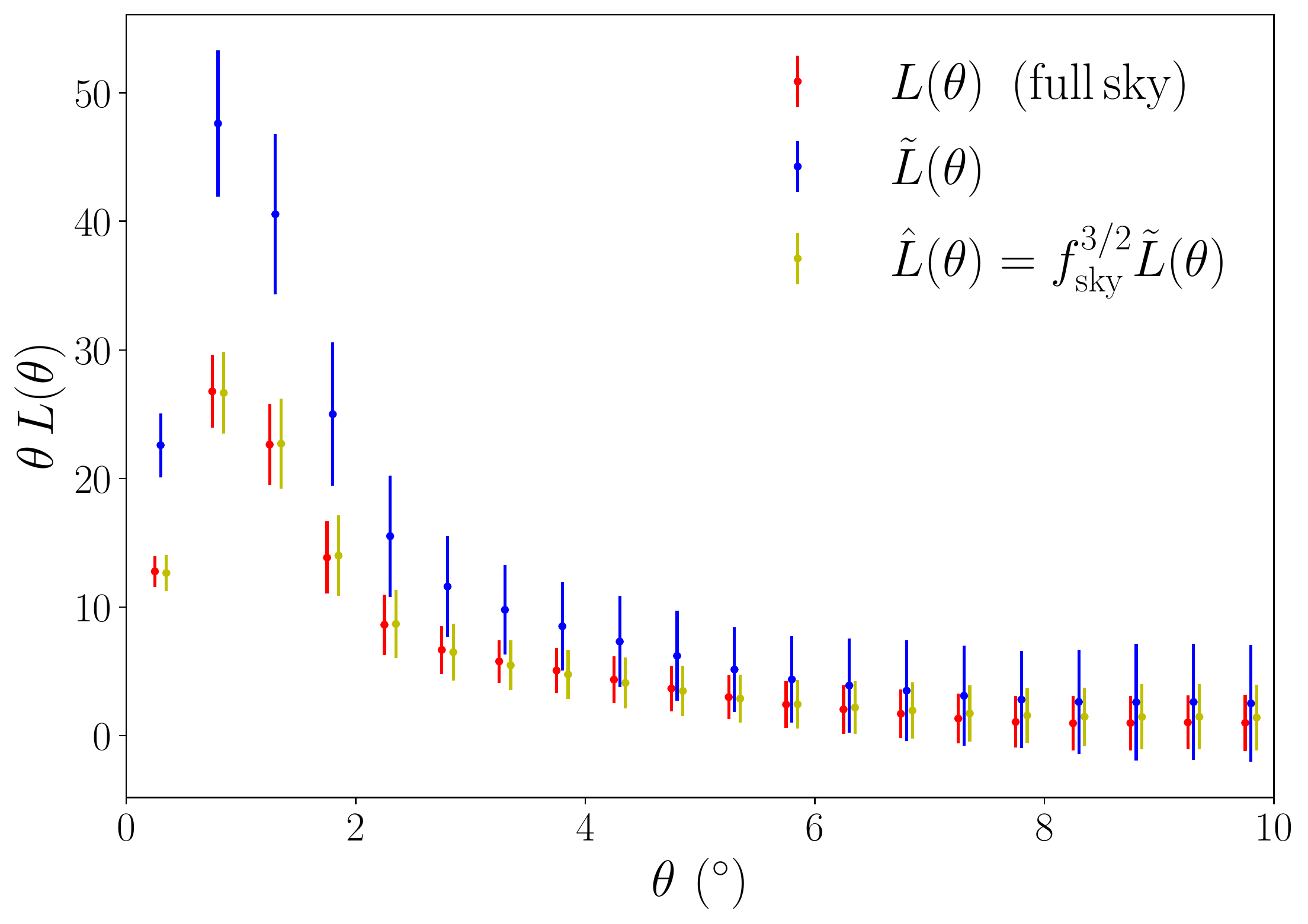}
      \caption{Projected LCF measured from an ensemble of 100 {\tt CoLoRe} simulations (mean and standard deviation). Results shown for measurements on the full sky (red), on a masked sky (blue), and correcting the latter by a factor $f_{\rm sky}^{3/2}$ as justified in the text.}\label{fig:fsky}
    \end{figure}

    To validate the simple estimator in Eq. \ref{eq:fsky} we run 100 fast mock realizations making use of {\tt CoLoRe} \cite{2111.05069}. We use first-order Lagrangian perturbation theory to generate a three-dimensional matter field in the lightcone, using a box of size $L_{\rm box}=870\,{\rm Mpc}\,h^{-1}$, enough to encompass the volume covered by the sample used here. The overdensity field was generated on a $1024^3$ grid, and an exponential bias model was used to match the large-scale bias of the 2MPZ sample. This field was then Poisson-sampled to match the number density and redshift distribution of 2MPZ. Each catalog thus generated was then masked as done for the real data, and the LCF was estimated using the method we just described. The LCF was then re-estimated using the full-sky catalog (thus requiring no correction). Fig. \ref{fig:fsky} shows the mean and scatter of the 100 masked LPT simulations before the $f_{\rm sky}^{3/2}$ correction in blue, the corrected LCF in yellow, and the mean of the unmasked simulations in red. We find that the $f_{\rm sky}^{3/2}$ correction is able to account for the effects of sky incompleteness almost exactly. This will in general depend on the mask complexity and on the small-scale correlations of the field, and thus the validity of this estimator should be tested before applying it to other observations.

  \section{Simulation-based emulator}\label{ssec:meth.emu}
    To build a theoretical prediction for the measured $\lcf(\theta)$, we use the \textsc{AbacusSummit} suite of high-performance cosmological $N$-body simulations \citep{Maksimova+2021}, designed to meet the simulation Requirements of the Dark Energy Spectroscopic Instrument (DESI) survey, and run with the high-accuracy cosmological code \textsc{Abacus} \citep{2019MNRAS.485.3370G,Garrison+2021}.
    We use the high-resolution \textsc{AbacusSummit} box, \texttt{AbacusSummit\_high\_c000\_ph100}, at the fiducial cosmology: $\Omega_b h^2 = 0.02237$, $\Omega_c h^2 = 0.12$, $h = 0.6736$, $10^9 A_s = 2.0830$, $n_s = 0.9649$. Its box size is $1000 \ {\rm Mpc}/h$, and it contains 6300$^3$ particles with mass $M_{\rm part} = 3.48 \times 10^{8} \ M_\odot/h$. 
    \begin{figure*}
      \centering
      \includegraphics[width=0.9\textwidth]{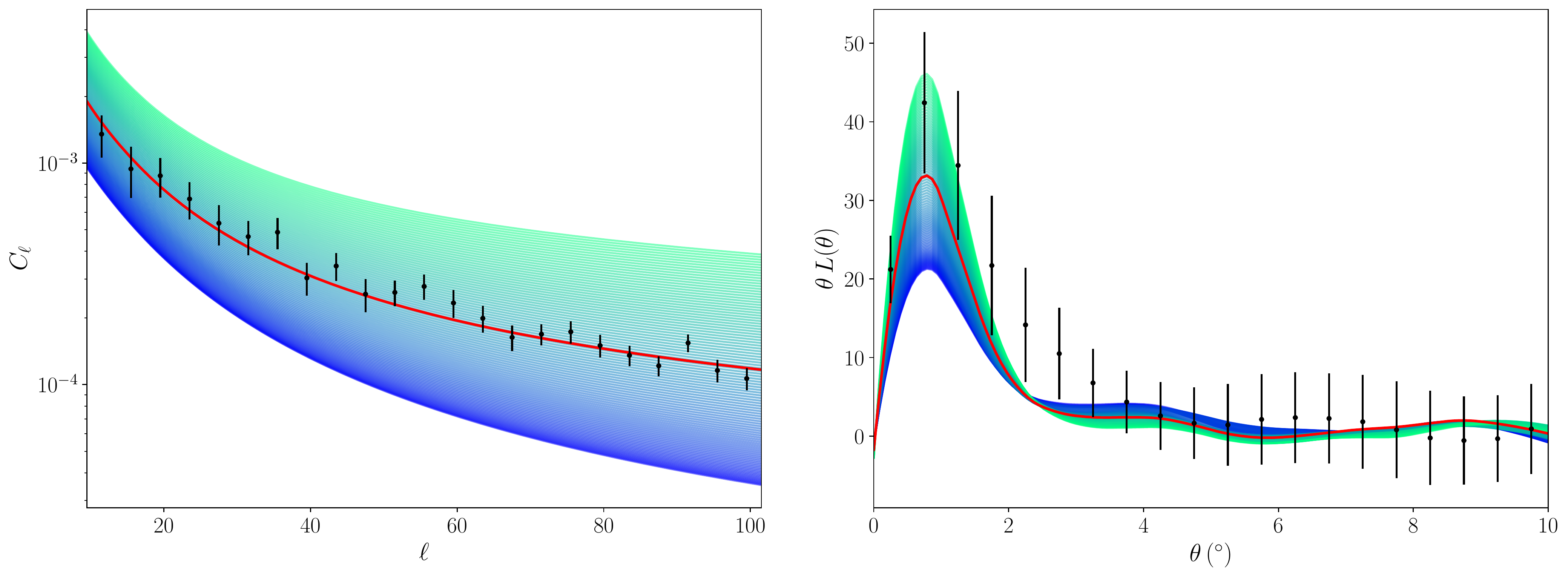}
      \caption{Power spectrum (left panel) and projected LCF (right panel) in the 2MPZ sample. The black dots with error bars show the measurements from the data. The red solid line shows the joint best-fit model. The coloured band shows the predictions for $\lmone=12.5$ while varying $\lmmin$ in the range $(10.7,12.2)$ (from dark blue to turquoise).}\label{fig:res_main}
    \end{figure*}
    
    We construct mock galaxy catalogues by applying the \textsc{AbacusHOD} model \citep{Yuan+2021} to the halo catalogue outputs at $z = 0.1$, generated using the halo-finding algorithm CompaSO \citep{2022MNRAS.509..501H}. The \textsc{AbacusHOD} model builds upon the baseline halo occupation distribution (HOD) model by incorporating various generalizations pertaining to halo-scale physics and assembly bias.
    In this work, we assume that the 2MASS sample is best approximated by so-called ``baseline HOD'' model with no decorations. The model is akin to the 5-parameter model of \cite{2007bZheng}, which gives the mean expected number of central and satellite galaxies per halo given halo mass $M$:
    \begin{align}
        \bar{N}_{\mathrm{cent}}^{\mathrm{LRG}}(M) & = \frac{1}{2}\mathrm{erfc} \left[\frac{\log_{10}(M_{\mathrm{min}}/M)}{\sqrt{2}\sigma_M}\right], \label{equ:zheng_hod_cent}\\
        \bar{N}_{\mathrm{sat}}^{\mathrm{LRG}}(M) & = \left[\frac{M-\kappa M_{\mathrm{min}}}{M_1}\right]^{\alpha}\bar{N}_{\mathrm{cent}}^{\mathrm{LRG}}(M).
        \label{equ:zheng_hod_sat}
    \end{align}
    Here, $M_{\mathrm{min}}$ characterizes the minimum halo mass to host a central galaxy. $M_1$ characterizes the typical halo mass that hosts one satellite galaxy. $\sigma_M$ describes the steepness of the transition from 0 to 1 in the number of central galaxies. $\alpha$ is the power law index on the number of satellite galaxies. $\kappa M_\mathrm{min}$ gives the minimum halo mass to host a satellite galaxy.

    Using this method, we generate simulated galaxy catalogs for 100 different HOD models, covering the range $\lmmin\in(10.7,12.2)$ and $\lmone\in(11.5,14.0)$ in a $10\times10$ rectangular grid. We set $\alpha = 1$, $\kappa = 1$ and $\sigma_M = 0.28$. We down-sample each catalog to match the redshift distribution of 2MPZ, and measure the angular power spectrum and the projected LCF of the resulting sample. To reduce the statistical noise in the power spectrum, we fit the measured $C_\ell$s to a 5$^{\rm th}$-order polynomial in log-log scale. We explored the possibility of correcting the statistical noise in the LCF measurements via principal component analysis (PCA), finding that our results were largely insensitive to the number of principal components kept. We thus do not apply any noise correction to the measured LCFs. The $C_\ell$ and $\lcf(\theta)$ were calculated with the same procedure used for the real data, including the bins of $\ell$ and $\theta$ described in the next section.

    From the measurements of $C_\ell$ and $\lcf(\theta)$ at the rectangular grid nodes, we build a simple emulator consisting of a set of bi-cubic interpolators in the $(\lmmin,\lmone)$ plane, one for each fixed scale ($\ell$ or $\theta$). Fig. \ref{fig:res_main} shows the theory predictions for both observables for a fixed $\lmone=12.5$, and $\lmmin$ in the range $(10.7,12.2)$, together with the measurements described in the next section.

  \section{Measurement of $\lcf(\theta)$ in 2MPZ}\label{ssec:result.lcf}
    We applied the estimator in Eq. \ref{eq:fsky} to the real 2MPZ catalog. A smoothing scale of $\sigma_\theta=1^\circ$ was used, corresponding to a comoving scale of $\sim6\,{\rm Mpc}$. We considered separation angles in the range $0<\theta<20^\circ$, in bins of width $\Delta\theta=0.5^\circ$. The resulting measurement is shown as black dots in the right panel of Fig. \ref{fig:res_main}.

    To quantify the additional constraining power brought about by the LCF, we also estimated the angular power spectrum of the sample, employing a pseudo-$C_\ell$ estimator as implemented in {\tt NaMaster} \cite{1809.09603}. We used bandpowers of width $\Delta\ell=4$, and kept only multipoles $\ell\leq100$, corresponding to a physical comoving scale of $k\sim0.5\,h\,{\rm Mpc}^{-1}$. The result is shown as black dots in the left panel of Fig. \ref{fig:res_main}.

    We estimated the covariance matrix of these measurements using the jackknife resampling method, dividing the footprint into $N_{JK}=400$ regions of equal area defined via $k$-means clustering \footnote{We used the public code \url{https://github.com/esheldon/kmeans_radec}.}.
    As in the case of most real-space correlations, the LCF covariance has large off-diagonal elements that must be taken into account. Additionally, although at tree level the 2PCF should be uncorrelated with the LCF, the data exhibit a non-negligible covariance between both measurements, sourced by higher-order connected $N$-point correlators.

    We estimate the statistical significance of this first measurement of the LCF on real data as $S/N=\sqrt{\chi^2-N_\theta}$, where $\chi^2\equiv\sum_{ij}\hat{\lcf}(\theta_i){\rm Cov}_{ij}^{-1}\hat{\lcf}(\theta_j)$ is the $\chi^2$ statistic of the measurement assuming no signal, and $N_\theta=20$ is the number of bins on which $\lcf(\theta)$ was calculated (i.e. the expectation value of $\chi^2$ in the absence of a signal). Doing so we obtain a $S/N=6.5$-$\sigma$ detection of phase correlations in 2MPZ. This significance is dominated by scales $\theta\lesssim4^\circ$ (corresponding to a physical separation of $\sim24\,{\rm Mpc}$), and depends mildly on the smoothing scale used. A larger smoothing of $\sigma_\theta=1.4^\circ$ results in a reduced significance of $\sim5.8\sigma$.

  \section{HOD constraints}\label{ssec:result.const}
    To quantify the additional information contained by the LCF (beyond that carried by the power spectrum), we use the simulation-based emulator described above to place constraints on the two free HOD parameters from our measurements of $\lcf(\theta)$ and $C_\ell$. To do so, we construct a Gaussian likelihood of the form
    \begin{equation}
      -2\log p({\bf d}|{\bf q})=({\bf d}-{\bf m}({\bf q}))^T\,{\sf C}^{-1}\,({\bf d}-{\bf m}({\bf q})),
    \end{equation}
    where ${\bf d}$ is the data, ${\sf C}$ is the jackknife covariance matrix, ${\bf q}\equiv(\lmmin,\lmone)$ are the free model parameters, and ${\bf m}({\bf q})$ is the theoretical model, provided by the emulator. In our case, the data ${\bf d}$ consists of the $C_\ell$ measurements, the $\lcf(\theta)$ measurements, or both combined. Since we assume flat priors on ${\bf q}$, given by the parameter range used to generate the emulator, the posterior distribution of the model parameters is simply proportional to the likelihood. The model is extremely fast, and we use a brute-force exploration of the posterior in a $100\times100$ grid.
\begin{figure}
  \centering
  \includegraphics[width=0.47\textwidth]{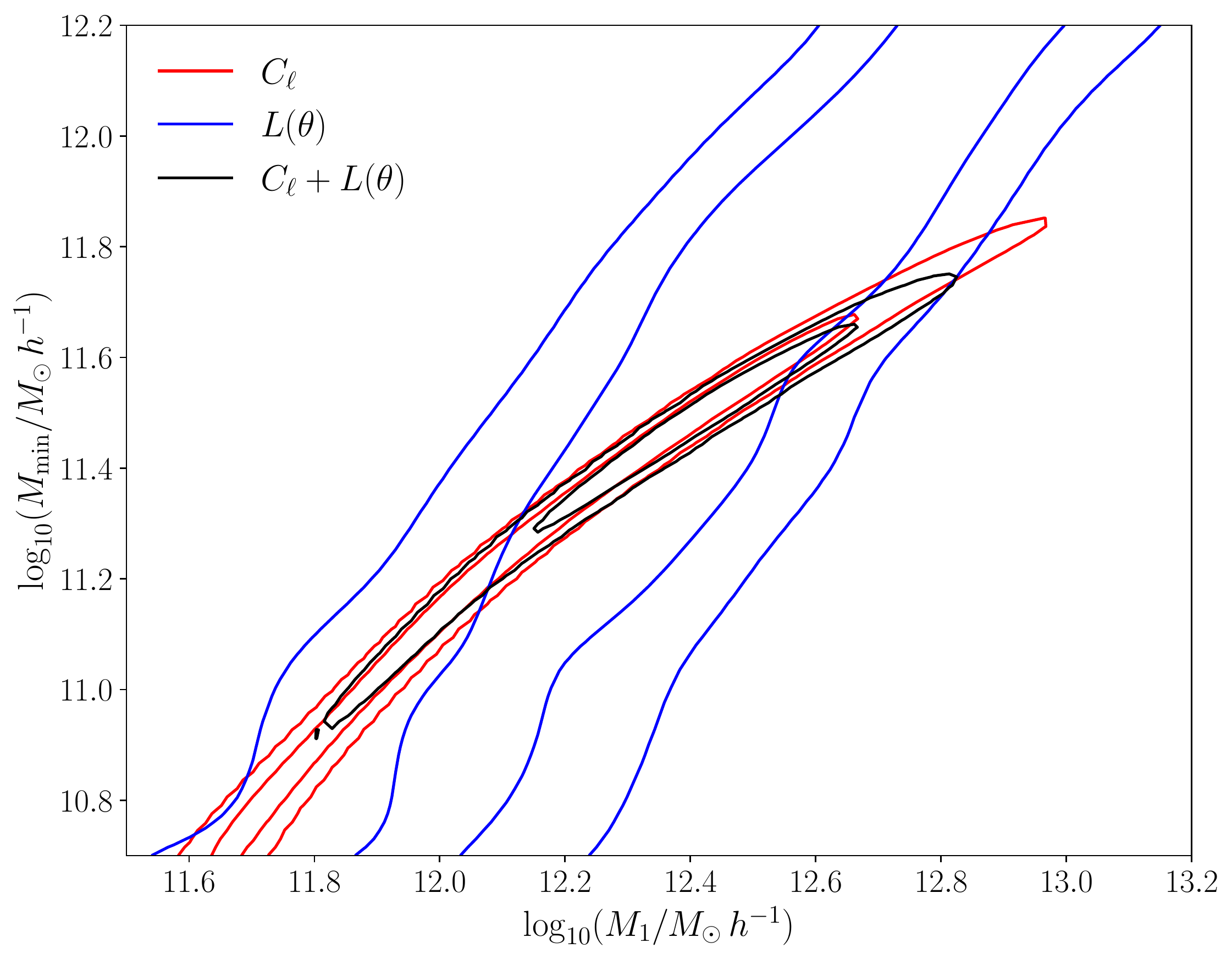}
  \caption{Constraints on the two HOD parameters $\lmmin$ and $\lmone$. The red and blue contours show the constraints using the power spectrum and the projected LCF respectively, while the joint constraints are shown in black. The combination is able to partially break the degeneracy between these parameters, significantly narrowing down the range of favoured characteristic halo masses.}\label{fig:res_const}
\end{figure}

    The resulting parameter constraints are shown in Fig. \ref{fig:res_const}. The red contours show the $68\%$ and $95\%$ confidence-level intervals obtained from the power spectrum alone. There is a strong degeneracy between both parameters that the $C_\ell$ alone is not able to break. The degeneracy direction roughly coincides with the ``constant-bias'' iso-contour, corresponding to those HOD models that yield the same large-scale galaxy bias. It is worth noting that the shot-noise contribution to the power spectrum was not subtracted from either the data or the emulator. This has a small effect on the scale dependence of the power spectrum that partially lifts this degeneracy, allowing us to discard models that lead to an exceedingly low galaxy number density. Nevertheless, at least within the scales used here, the $C_\ell$ alone would only allow us to place an upper bound on the HOD mass scales.

    The blue contours in Fig. \ref{fig:res_const} show the constraints derived from the $\lcf(\theta)$ measurements. Unsurprisingly, on its own, the LCF is unable to constrain the HOD masses, which are broadly degenerate along the $\lmmin\sim1.7\lmone-9.6$ direction. Nevertheless, this degeneracy direction, is different from that of the power spectrum, and thus the combination of both measurements is able to significantly improve the final constraints. The results for the combination of $C_\ell$ and $\lcf(\theta)$ are shown in black in the same figure. The addition of the LCF places a lower bound on the HOD masses, and we obtain the marginalized constraints (mean and standard deviation):
    \begin{eqnarray}
        \lmmin & = & 11.45\pm0.17,\nonumber \\
        \lmone & = & 12.37\pm0.20.
    \end{eqnarray}
    The joint posterior has a peak at $\lmmin=11.50$, $\lmone=12.42$. This model is a good fit to the joint data vector, with $\chi^2/{\rm d.o.f.}=0.73$ for 41 degrees of freedom (corresponding to a probability to exceed ${\rm PTE}=0.83$). The median halo mass for this best-fit model is $M_{\rm median}\simeq1.4\times10^{12}\,M_\odot$. The model also predicts a relatively high abundance of satellites, with an average of  $0.77$ satellites per central.

\section{Conclusions}
  We have presented the case of projected phase correlations as a probe of the non-Gaussian features in the late-time matter distribution, focusing on a specific configuration: the line correlation function. We have shown that, for a sufficiently simple sky mask, and at least in the case of galaxy clustering, a simple scaling by $f_{\rm sky}^{3/2}$ is sufficient to recover an unbiased measurement of phase correlations. We have then applied this simple estimator to data from the 2MPZ survey, which is particularly well suited for a first measurement of this observable. Thus we have presented the first measurement to date of phase correlations on real data, which we detect at $6.5\sigma$. 

  We have made use of the \textsc{AbacusSummit} suite of simulations to create a basic emulator of the LCF and the power spectrum as a function of two HOD parameters: $\lmmin$, the minimum mass to form central galaxies, and $\lmone$, the mass scale for satellites. Applying this emulator to the measurements of $\lcf(\theta)$ and $C_\ell$ in 2MPZ, we have then shown that the addition of the LCF is able to significantly improve the parameter constraints achievable with the power spectrum alone. In particular, we have shown that the combination of both observables is able to break the degeneracy between the two HOD masses, allowing us to place reasonably tight constraints on both.

  This opens the door to using phase correlations in cosmological data analysis, to unlock additional information contained in the non-Gaussian signatures of the large-scale structure. A broader application of the LCF in cosmology will need to tackle a number of outstanding challenges, however. With the increased statistical power of larger samples, more sophisticated estimators accounting for the effects of survey geometry will likely be necessary, or the development of methods to forward-model them in the theory prediction. More accurate theoretical predictions will also be necessary. This is generally challenging for non-Gaussian probes, often requiring the use of simulation-based emulators. Although this has been our approach in this paper, it may be possible to develop sufficiently accurate analytical predictions based on perturbation theory (see \cite{Oliveira_inprep}). In any case, the regime of reliability of these predictions will need to be carefully calibrated, especially in the context of constraints on fundamental cosmological parameters for which phase correlations could be particularly powerful. Finally, although we have focused only on the LCF, other phase correlation configurations exist which may contain valuable information. As in the case of the bispectrum, identifying the most relevant configuration will be of vital importance in order to focus future data analysis efforts.

  Regardless of these challenges, exploiting the information encoded in non-Gaussian probes of the late-time matter fluctuations, such as the LCF, is a vital endeavour to make the most of near-future Stage-IV surveys.

\begin{acknowledgments}
  We thank Raul Angulo, Harry Desmond, Joyce Byun and Sebastian von Hausseger for useful comments and discussions. FOF acknowledges support from Swiss National Science Foundation through an Early Postdoc Mobility fellowship, grant reference P2GEP2\_199989. DA is supported by the Science and Technology Facilities Council through an Ernest Rutherford Fellowship, grant reference ST/P004474. This research used resources of the National Energy Research Scientific Computing Center (NERSC), a U.S. Department of Energy Office of Science User Facility located at Lawrence Berkeley National Laboratory. We also made extensive use of computational resources at the University of Oxford Department of Physics, funded by the John Fell Oxford University Press Research Fund.
\end{acknowledgments}

\bibliography{main,extra}

\end{document}